\documentclass[secnumarabic,amssymb, nobibnotes, aps, prd, 11pt, abstract]{revtex4-2}
\usepackage{tocloft}
\usepackage[dvips]{graphicx}
\usepackage{amsmath}
\usepackage{subcaption}
\usepackage{hyperref}
\usepackage{times}
\usepackage{filecontents}
\hypersetup{colorlinks,linkcolor={red},citecolor={blue},urlcolor={red}}
\usepackage{setspace}
\usepackage{xcolor}
\pagecolor{white}

\usepackage{multirow}
\color{black}

\begin{document}

\singlespacing

	\title{Nonperturbative Isentropic Processes in AdS Black Holes with Nonlinear Electrodynamics}%

	\author{Mozib Bin Awal$^1$}
	
	\email{$rs_mozibbinawal@dibru.ac.in$}
	
	\author{Prabwal Phukon$^{1,2}$}
	\email{$prabwal@dibru.ac.in$}
	
	\affiliation{$^1$Department of Physics, Dibrugarh University, Dibrugarh, Assam,786004.\\$^2$Theoretical Physics Division, Centre for Atmospheric Studies, Dibrugarh University, Dibrugarh,Assam,786004.\\}

	\begin{abstract}
	
	We study the isentropic processes in a class of Anti de Sitter black holes coupled to non-linear electrodynamics. We demonstrate that such processes are classically forbidden but can proceed via quantum mechanical tunnelling. We compute the Euclidean action associated with the tunnelling process and analyze its dependence on the black hole charge, horizon radius, and the non-linear electrodynamics parameters characterizing each model. We find that the tunnelling probability is increasingly suppressed as the strength of the non-linearity is enhanced. We further find that smaller black holes exhibit a significantly higher tunnelling probability compared to larger ones, indicating a departure from classical behaviour. We conjecture that this behaviour may be universal across a broad class of black hole spacetimes. We discuss the implications of our results for entropy bounds and their potential relevance to the black hole information loss paradox.

\end{abstract}
	
	\maketitle
	
\section{Introduction}\label{sec1}
Hawking, in his celebrated paper, showed that a black hole can eventually evaporate by a process what we call the Hawking radiation \cite{Hawking}. This naturally raises a fundamental question of whether once the black hole has vanished, is it possible to reconstruct the information associated with the matter that collapsed to form it? One of the most persistent open problems in theoretical physics concerns the fate of information in black hole spacetimes \cite{Hawking2}. If one assumes that information is encoded in Hawking radiation within a semiclassical description, internal inconsistencies arise because either quantum information must be duplicated \cite{Yeom}, violating the no-cloning principle, or the standard relations governing entanglement entropy cease to hold \cite{Almheiri:2012rt}. Conversely, abandoning information conservation altogether conflicts not only with holographic principles \cite{Maldacena:1997re} but also with the fundamental quantum framework that underlies the mechanism of black hole evaporation itself \cite{Banks:1983by}. A close analysis of the black hole information paradox reveals that it rests on several fundamental assumptions \cite{Ong:2016iwi}. First it is assumed that the complete evolution of a black hole, from its formation to its final evaporation, is governed by unitary quantum dynamics. Secondly, it is assumed that the properties of Hawking radiation can be fully described using local quantum field theory on a curved spacetime background. Third, general covariance is presumed to hold throughout spacetime, with the possible exception of regions containing curvature singularities. Fourth, the identification of black hole entropy with the area of the event horizon, as expressed by the Bekenstein-Hawking relation \cite{Bekenstein}, or more generally by the Iyer-Wald entropy in alternative theories of gravity \cite{Iyer}. Finally, it is assumed that there exists a hypothetical observer capable of accessing and measuring the information or entanglement encoded in the emitted Hawking radiation. The information paradox arises because these assumptions cannot all hold simultaneously. Although no definitive resolution to this paradox is currently known, it is generally expected that if information is preserved during black hole evaporation, then the associated dynamics must involve effects beyond the semiclassical description of spacetime and involve intrinsically nonperturbative quantum gravitational effects.

Starting in the 1970s, a number of universal bounds on entropy were independently proposed for general physical systems. One of the earliest and most influential among these is the Bekenstein bound, which was originally derived using semiclassical considerations. It asserts that for a system with finite energy confined to a finite region of space, the entropy cannot exceed a certain upper limit \cite{Bekenstein:1974ax,Bekenstein:1973ur,Bekenstein:1972tm,Bekenstein:1980jp}. Over the years, the Bekenstein bound has been widely studied and generalized, with its implications examined across a broad range of physical contexts. For example, Ref. \cite{bound1} investigated the implications of the bound for the information-carrying capacity of quantum communication channels, Ref. \cite{bound2} tested its applicability to bound states in interacting field theories and demonstrated that the bound continues to hold when interaction effects are consistently taken into account. From a holographic viewpoint, Ref. \cite{bound3} argued that the Bekenstein bound remains valid across all regimes of gravitational coupling, provided certain conditions are satisfied. In addition, numerous extensions and alternative formulations of entropy bounds have been developed, such as the causal entropy bound \cite{bound4}, the Unruh–Wald bound \cite{bound5,bound6} , various generalized versions of the Bekenstein bound \cite{bound7}, as well as covariant entropy bounds \cite{bound8,bound9} and their further generalizations \cite{bound10,bound11}.

In addition to the various formulations discussed above, the Bekenstein bound has been interpreted from several complementary perspectives. For example, Ref. \cite{bound12} proposed that the bound may be understood as a consequence of the Pauli exclusion principle, assuming that the fundamental microscopic constituents of a black hole are finite in number and fermionic in nature; related analyses of such fermionic degrees of freedom were presented in Refs. \cite{bound13,bound14}. Other investigations have focused on the role of the bound in information-theoretic processes. In particular, Ref. \cite{bound15} argued that the entropy bound regulates the redistribution of degrees of freedom in generic physical systems and leads to an effective loss of accessible information through entanglement between spacetime geometry and quantum fields. Furthermore, Ref. \cite{bound16} employed the bound to place limits on the information storage capacity of black hole remnants, thereby challenging remnant-based resolutions of the information loss paradox. Taken together, the arguments discussed above are directly relevant to the black hole information loss problem, as they highlight a fundamental tension among several basic principles discussed earlier.

It should be emphasized that the equivalence between Bekenstein-Hawking entropy and Boltzmann entropy arises within a semiclassical framework. However, very recently, in Ref \cite{Mann:2025ojd}, the authors have argued that nonperturbative effects can violate this equivalence, allowing the Boltzmann entropy to become larger than the Bekenstein–Hawking value. Specifically, Ref \cite{Mann:2025ojd} studied the isentropic absorption of charged particles by a Reissner-Nordström black hole within the WKB approximation and suggested that such processes may provide a nonperturbative mechanism through which entropy bounds can be exceeded. Following this work, the authors in \cite{Dubey:2025imq} showed that within any diffeomorphism-invariant theory of gravity, the isentropic capture of a classical charged test particle is prohibited at the classical level for all stationary, nonextremal, axisymmetric black holes in four spacetime dimensions. They further consider the Kerr-Newman black hole within general relativity and investigate, through a quantum tunneling framework, the circumstances under which isentropic absorption can occur. Motivated by these studies, in this paper, we consider four black hole systems sourced by nonlinear electrodynamics (NLED). By analyzing isentropic absorption processes in these NLED black hole backgrounds, we aim to examine whether the conclusions drawn for Reissner-Nordström and Kerr-Newman spacetimes persist in the presence of nonlinear electromagnetic interactions, and to assess the extent to which such interactions may influence entropy bounds beyond the semiclassical regime.

\section{Radial Geodesic Motion }\label{sec2}
In this section, we briefly revisit the derivation of the radial geodesic equation and the associated effective potential, following the analysis presented in Ref. \cite{Mann:2025ojd}. We begin by considering the most general line element describing a static, spherically symmetric spacetime
\begin{equation}\label{eq1}
ds^2=-g_{00}(r)dt^2+g_{11}(r)dr^2+r^2d\Omega^2
\end{equation}
We consider a particle with rest mass $m$ and charge $q$. Then the asymptotic energy of the particle will be 
\begin{equation}\label{eq2}
E=\gamma m\sqrt{g_{00}}+q\Phi
\end{equation}
where $\sqrt{g_{00}}$ is the redshift factor, $\gamma=1/\sqrt{1-v^2}$ is the Lorentz factor and $q\Phi$ is the electrostatic energy. Therefore, we can write,
\begin{equation}\label{eq3}
\gamma = \frac{1}{m\sqrt{g_{00}}} \left( E - q \Phi \right)
\end{equation}
hence we obtain,
\begin{equation}\label{eq4}
\gamma^{2} v^{2} = \gamma^{2} - 1 = \frac{1}{m^{2} g_{00}} \left( E - q \Phi \right)^{2} - 1
\end{equation}
For a particle undergoing purely radial motion (with radial distance $d\ell = \sqrt{g_{11}} dr$), with vanishing angular momentum, the rate of change of the radial coordinate with respect to its proper time is given by
\begin{equation}\label{eq5}
\frac{d\ell}{d\tau} = \pm \gamma v
\end{equation}
so, \begin{equation}\label{eq6}
\left( \frac{dr}{d\tau} \right)^{2} = \frac{\gamma^{2} v^{2}}{g_{11}} = \frac{1}{m^{2} g_{00} g_{11}} \left( E - q \Phi \right)^{2} - \frac{1}{g_{11}}\
\end{equation}
or equivalently, 
\begin{equation}\label{eq7}
\left( \frac{dr}{d\tau} \right)^{2}
+ V_{\mathrm{eff}} (r) = 0
\end{equation}
where $V_{\mathrm{eff}}$ is termed as the effective potential and is given by
\begin{equation}\label{eq8}
V_{\mathrm{eff}} (r) = \frac{\gamma^{2} v^{2}}{g_{11}} = -\frac{1}{m^{2} g_{00} g_{11}} \left( E - q \Phi \right)^{2} + \frac{1}{g_{11}}
\end{equation}
This form of the effective potential is valid for any theory of gravity and for the metric of the form
\begin{equation}\label{eq9}
 ds^{2} = - f(r) dt^{2} + \frac{1}{f(r)} dr^{2} + r^{2} d\Omega^{2}
\end{equation}
the effective potential becomes 
\begin{equation}\label{eq10}
V_{\mathrm{eff}} (r) = - \frac{1}{m^{2}} \left( E - q\Phi \right)^{2} + f(r)
\end{equation}

\section{Isentropic absorption of charged particle}\label{sec3}
For any isentropic process, the entropy remains constant and so we have,
\begin{equation}\label{eq11}
TdS=0=dM-\Phi dQ
\end{equation}
Interpreting the infinitesimal changes in the black hole mass and charge as the energy and charge of the absorbed particle,  $dM \rightarrow E$ and $dQ \rightarrow q$, the isentropic condition reduces to,
\begin{equation}\label{eq12}
E - q\Phi = 0
\end{equation}
In the following subsections, we demonstrate for a few NLED black hole systems that such an isentropic process is classically forbidden. We further show that quantum tunnelling effects can render such processes possible and examine how the corresponding tunnelling probabilities depend on the parameters characterizing the black hole.

\subsection{NED AdS Black Holes}
\begin{figure}[t]
\centering
\includegraphics[width=0.5\linewidth]{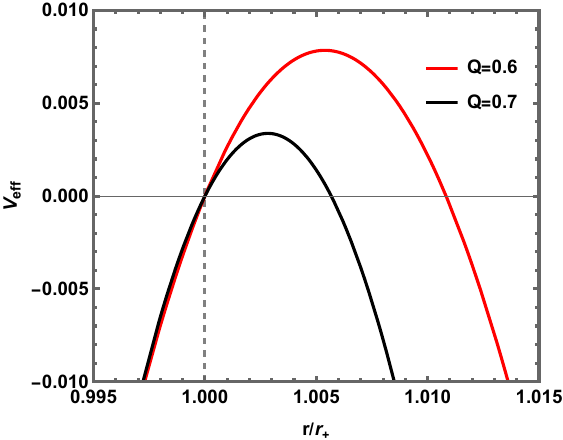}
\caption{Effective potential experienced by the charged test particle for an isentropic absorption by NED AdS black hole} 
\label{fig1}
\end{figure}

The first black hole system under consideration is the Non-linear Electrodynamics AdS black hole with the action \cite{Kruglov:2022mde}
\begin{equation}\label{eq13}
I=\int d^4x\sqrt{-g}\left(\frac{R-2\Lambda}{16\pi G}+\mathcal{L}(\mathcal{F})\right)
\end{equation}
with $\Lambda=-3/l^2$ being the cosmological constant, $l$ is the AdS length and $G$ is the Newton's constant. The Lagrangian of the NED-AdS black hole is expressed as \cite{Bron}
\begin{equation}\label{eq14}
\mathcal{L}(\mathcal{F})=-\frac{\mathcal{F}}{4\pi\cosh^2\left(a\sqrt[4]{2|\mathcal{F}|}\right)}
\end{equation}
where $a$ is termed as the coupling constant and $\mathcal{F}=F^{\mu\nu}F_{\mu\nu}/4$ is the field invariant. The line element is expressed as follows
\begin{equation}\label{eq15}
ds^2=-f(r)dt^2+\frac{1}{f(r)}dr^2+r^2(d\theta^2+\sin^2\theta d\phi^2)
\end{equation}
The lapse function in equation \ref{eq15} is expressed as
\begin{equation}\label{eq16}
f(r)=1-\frac{2MG}{r}+\frac{Q^2G}{br}\tanh\left(\frac{b}{r}\right)+\frac{r^2}{l^2}
\end{equation}
$b=a\sqrt{Q}$ is used here for simplicity. We write down the effective potential experienced by a charged test particle having charge $q$ in the NED AdS background using equation \ref{eq10}
\begin{equation}\label{eq17}
 V_{\mathrm{eff}}(r)=-\frac{1}{m^2}\left[q\frac{Q \tanh \left(\frac{b}{r_+}\right)}{b}-q\frac{Q \tanh \left(\frac{b}{r}\right)}{b}\right]^2+f(r)
\end{equation}
Now, we have $V_{\mathrm{eff}}(r_+)=0$ and $\left.\frac{dV_{\mathrm{eff}}}{dr}\right|_{r_+}=4\pi T$. Performing a near horizon expansion we obtain 
\begin{equation}
V_{\text{eff}}(r)=V_{\text{eff}}(r_{+})+4\pi T (r - r_{+})+\mathcal{O}\!\left( (r - r_{+})^{2} \right)
\end{equation}
Therefore, in the region just outside the event horizon, $r>r_+$, the effective potential remains positive ($V_{\text{eff}}>0$). As a consequence of the radial equation of motion ( Eq. \ref{eq7}), the particle cannot propagate in this region, implying that the absorption process is classically prohibited. We show the plot of the effective potential against $r/r_+$ in Figure \ref{fig1}. The interval extending from the event horizon ($r=r_+$) to the radial location at which the effective potential vanishes again (denoted $r=r_2$) constitutes a classically forbidden region. Within this range, the particle is unable to traverse the barrier through classical motion and therefore cannot cross $r_2$.
We also observe that increasing the charge decreases the height as well as the width of the effective potential.

Within the path-integral formulation, the quantum transition amplitude associated with the radial degree of freedom $r(\lambda)$ can be expressed as a sum over all possible trajectories propagating in an effective potential $V_{\text{eff}}$. Each trajectory contributes with a weight determined by the classical action, such that the amplitude takes the schematic form 
\begin{equation}\label{eq19}
K \; \propto \; \int \mathcal D r \; \exp\!\left(\tfrac{i}{\hbar} S\right),
\end{equation}
where $S$ denotes the action functional evaluated along a given path. Quantum evolution is therefore obtained by coherently summing over all admissible histories, with the phase of each contribution governed by the action. In regions where classical motion is allowed, the dominant contributions arise from paths near the classical solution. However, in classically forbidden domains, characterized by a positive effective potential, the path integral is instead controlled by non-classical trajectories. In this case, it is convenient to perform an analytic continuation to imaginary (Euclidean) time, implemented through a Wick rotation of the affine parameter, $\lambda \to -i\lambda_E$. Under this transformation, the action becomes purely Euclidean, $S \to iS_E$ and the oscillatory phase factor is converted into an exponentially suppressed weight. As a result, the transition amplitude acquires the form
\begin{equation}\label{eq20}
K \; \sim\; e^{-S_E/\hbar}, 
\qquad 
S_E = \int d\lambda_E \, L_E\,,
\end{equation}
where the Euclidean action is given by
\begin{equation}\label{eq21}
S_E=\int d\lambda_EL_E
\end{equation}
Here $L_E$ denotes the Lagrangian after continuation to Euclidean time. In the semiclassical limit, the leading contribution to the path integral arises from the Euclidean classical path, namely the solution of the Euler–Lagrange equations defined in imaginary time that interpolates between the classical turning points of the effective potential. Because the transition amplitude is suppressed by a factor of the form $e^{-S_E/\hbar}$, the corresponding tunnelling probability is obtained by taking the squared magnitude of this amplitude, yielding
\begin{equation}\label{eq22}
\Gamma \;\propto\; |K|^2 \;\sim\; e^{-2S_E/\hbar}.
\end{equation}
Adopting units in which $\hbar=1$ the probability for the particle to traverse the classically forbidden interval $[r_{2}, r_{+}]$ may therefore be written as \cite{Mann:2025ojd}
\begin{equation} \label{eq23}
    \Gamma \;\approx\; e^{-2S_E}.
\end{equation}
The Euclidean action $S_E$ is explicitly given by 
\begin{align}
    S_E &= \int _{r_2} ^{r_+} L_E d\lambda_E \\
    & = \int _{r_2} ^{r_+} \frac{1}{dr/d\lambda_E}d r \\
    & = - \int _{r_2} ^{r_+} \frac{1}{\sqrt{V_{\text{eff}}(r)}} dr \\  & = \int _{r_+} ^{r_2} \frac{1}{\sqrt{V_{\text{eff}}(r)}} dr  \label{eq27}.
\end{align}

\begin{figure}[h!]
    \centering
    \begin{subfigure}[b]{0.45\textwidth}
        \centering
        \includegraphics[width=\textwidth]{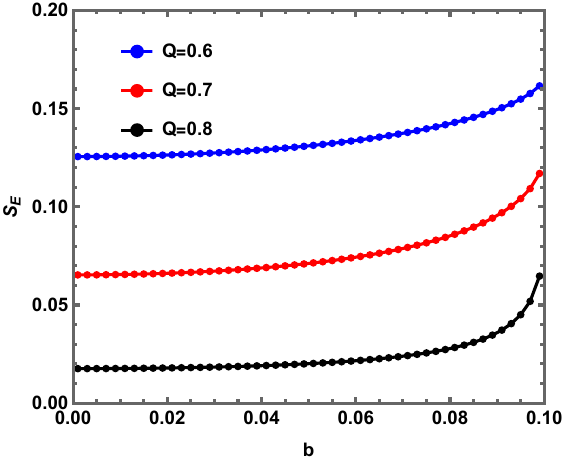}
        \caption{$S_E$ versus $b$ for fixed $Q$}
        \label{f2a}
    \end{subfigure}
    \hspace{0.03\textwidth}
    \begin{subfigure}[b]{0.45\textwidth}
        \centering
        \includegraphics[width=\textwidth]{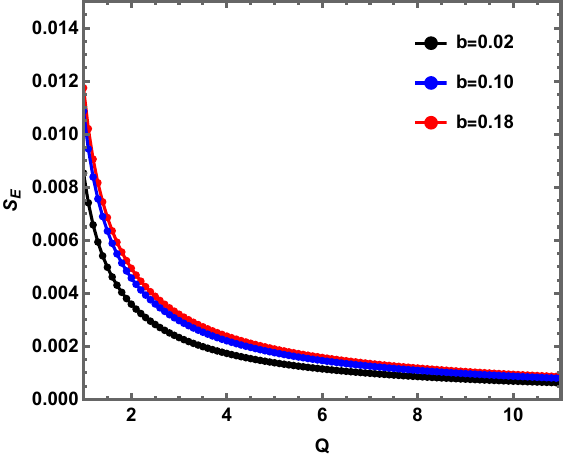}
        \caption{$S_E$ versus $Q$ for fixed $b$}
        \label{f2b}
    \end{subfigure}

    \vskip\baselineskip  

    \begin{subfigure}[b]{0.45\textwidth}
        \centering
        \includegraphics[width=\textwidth]{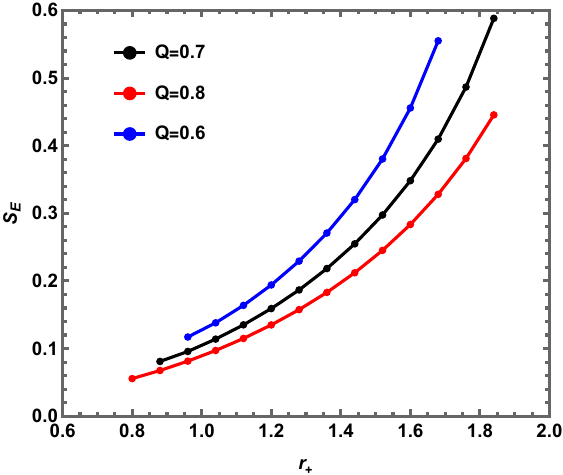}
        \caption{$S_E$ versus $r_+$ for fixed $b$}
        \label{f2c}
    \end{subfigure}
    \hspace{0.03\textwidth}
    \begin{subfigure}[b]{0.45\textwidth}
        \centering
        \includegraphics[width=\textwidth]{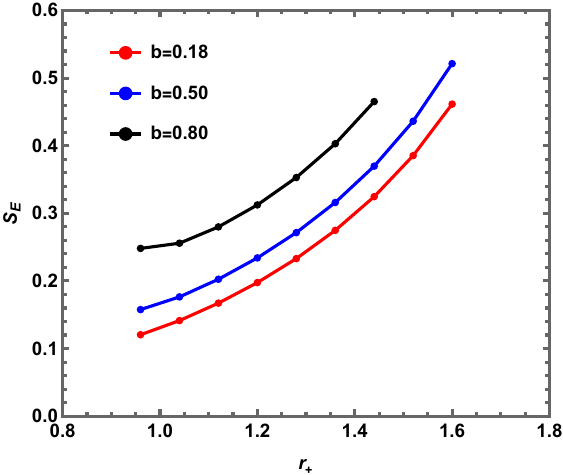}
        \caption{$S_E$ versus $r_+$ for fixed $Q$}
        \label{f2d}
    \end{subfigure}

    \caption{The plots presented above depict the Euclidean action associated with the isentropic absorption of a charged particle by NED-AdS black hole. Here we considered $m=0.0002$, $q=0.005$}
    \label{f2}
\end{figure}

The integral given in equation \ref{eq27} is extremely difficult to solve analytically. So instead we solve it numerically and plot the Euclidean action $S_E$ for real values of turning points $r_+$ and $r_2$. From the relationship between the tunnelling probability and the Euclidean action as given by equation \ref{eq23}, it follows that larger values of the action lead to stronger exponential suppression, while smaller values enhance the probability. Consequently, Figure \ref{f2} can be directly interpreted as illustrating the behaviour of the tunnelling probability within the semiclassical approximation employed here. We provide the plots of the Euclidean action with the non-linear parameter $b$ in Figure \ref{f2a}, with charge in Figure \ref{f2b} and with the horizon radius in Figure \ref{f2c} and \ref{f2d}. 

We observe that the Euclidean action initially exhibits an approximately linear dependence on the nonlinear parameter $b$ and after a certain value of $b$ is reached, the action increases rapidly and almost exponentially. This qualitative behaviour is consistently observed for all fixed values of the black hole charge $Q$ as shown in Figure \ref{f2a}. In terms of the tunnelling probability, this implies that the tunnelling probability becomes increasingly suppressed once the nonlinear parameter exceeds a certain threshold. Consequently, for black holes governed by stronger nonlinear electrodynamics, an isentropic absorption of a charged test particle is progressively less probable. Moreover, we also observe from Figure \ref{f2a} that for a fixed $b$ increasing the charge of the black hole $Q$ decreases the Euclidean action, implying a larger corresponding tunnelling probability. 

On the other hand, the Euclidean action decreases exponentially as the charge of the black hole increases when the non linear parameter is kept fixed as depicted in Figure \ref{f2b}. Therefore, the corresponding tunnelling will increase exponentially as $Q$ keeps on increasing implying that an isentropic absorption gets more and more likely as the charge of the black hole increases. Also, for a fixed value of the charge parameter $Q$ increasing the non linear parameter $b$ increases the Euclidean action $S_E$ and hence suppressing the tunnelling probability. 

Finally, from Figure \ref{f2c} and \ref{f2d} we find that the Euclidean action increases non linearly as the size of the black hole increases characterised by the increasing horizon radius $r_+$. This behaviour remains consistent for both fixed $b$ and $Q$. It therefore follows that smaller black holes exhibit a higher tunnelling probability compared to their larger counterparts. This indicates that smaller black holes behave in a less classical manner, as they are more susceptible to isentropic absorption of charged particles and can more readily lead to violations of entropy bounds than larger black holes. In this case also, we observe that for a black hole of a particular horizon radius, increasing the charge decreases the Euclidean action, enhancing the tunnelling probability. Whereas, increasing the non linear parameter increases the the Euclidean action, suppressing the tunnelling probability.

\subsection{ModMax AdS Black Holes}
\begin{figure}[h!]
\centering
\includegraphics[width=0.5\linewidth]{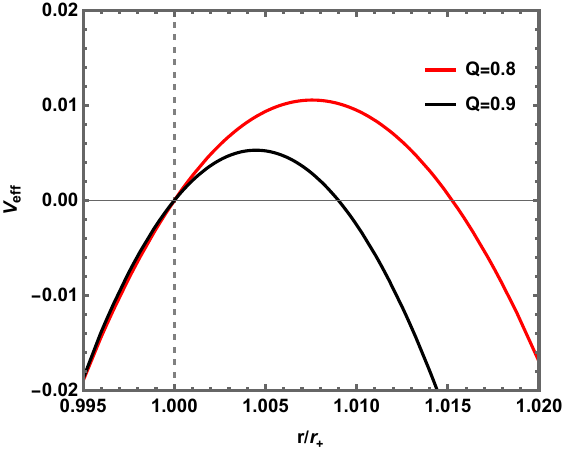}
\caption{Effective potential experienced by the charged test particle for an isentropic absorption by ModMax AdS black hole} 
\label{fig3}
\end{figure}

\begin{figure}[h!]
    \centering
    \begin{subfigure}[b]{0.45\textwidth}
        \centering
        \includegraphics[width=\textwidth]{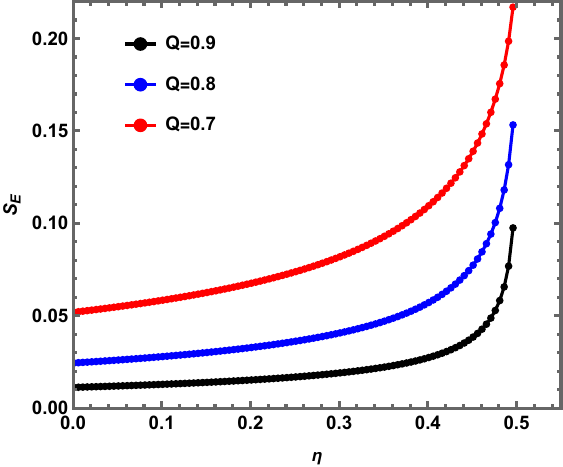}
        \caption{$S_E$ versus $\eta$ for fixed $Q$}
        \label{f4a}
    \end{subfigure}
    \hspace{0.03\textwidth}
    \begin{subfigure}[b]{0.45\textwidth}
        \centering
        \includegraphics[width=\textwidth]{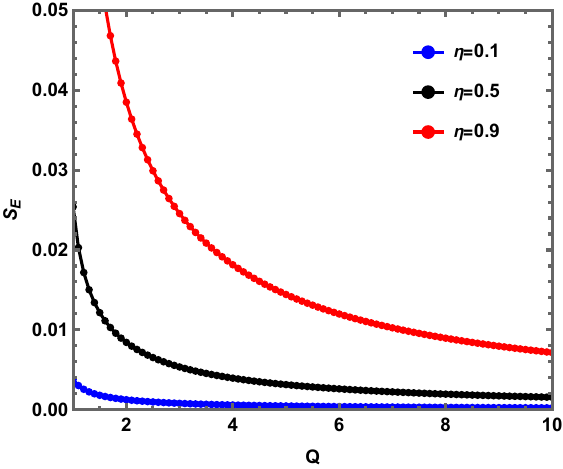}
        \caption{$S_E$ versus $Q$ for fixed $\eta$}
        \label{f4b}
    \end{subfigure}

    \vskip\baselineskip  

    \begin{subfigure}[b]{0.45\textwidth}
        \centering
        \includegraphics[width=\textwidth]{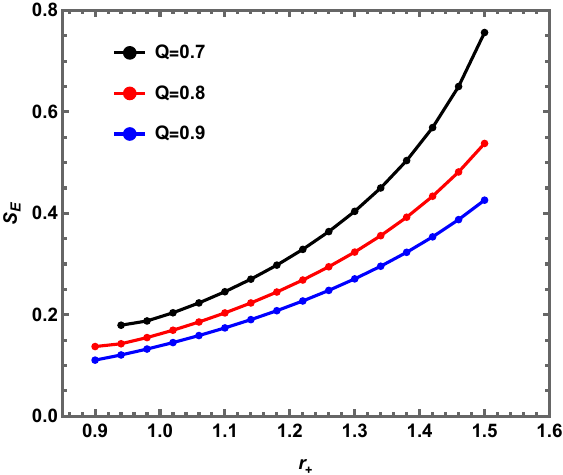}
        \caption{$S_E$ versus $r_+$ for fixed $\eta$}
        \label{f4c}
    \end{subfigure}
    \hspace{0.03\textwidth}
    \begin{subfigure}[b]{0.45\textwidth}
        \centering
        \includegraphics[width=\textwidth]{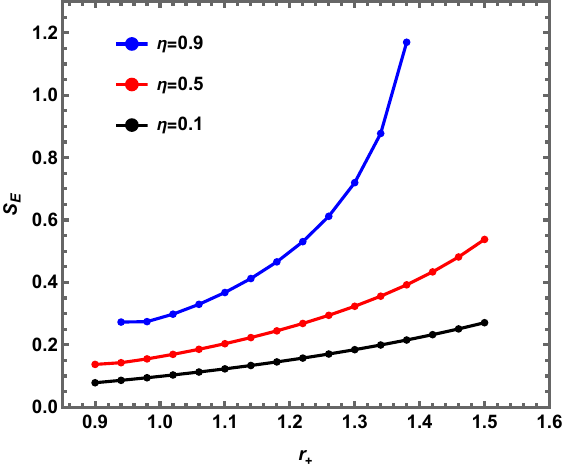}
        \caption{$S_E$ versus $r_+$ for fixed $Q$}
        \label{f4d}
    \end{subfigure}

    \caption{The plots presented above depict the Euclidean action associated with the isentropic absorption of a charged particle by ModMax-AdS black hole. Here we considered $m=0.0002$, $q=0.005$}
    \label{f4}
\end{figure}

The next black hole system we consider is the ModMax AdS black hole whose action is given by \cite{Sekhmani:2025kav}
 \begin{equation}\label{eq28}
\mathcal{I}=\frac{1}{16\pi}\int d^4x\sqrt{-g}\left(\mathcal{R}+6L^{-2}-4\mathcal{L}\right)
\end{equation}
Here $\mathcal{R}$ denotes the Ricci scalar, $L$ represents the AdS length scale, $g$ is the determinant of the spacetime metric and $\mathcal{L}$ denotes the ModMax Lagrangian, which can be written in the following form  \cite{Sekhmani:2025kav}
\begin{equation}\label{eq29}
\mathcal{L}=\frac{1}{2}\left(\mathcal{S}\cosh{\eta}-\sqrt{\mathcal{S}^2+\mathcal{P}^2}\sinh{\eta}\right)
\end{equation}
where the parameter $\eta$ is a dimensionless intrinsic quantity characterizing the ModMax theory, and $\mathcal{S}$ and $\mathcal{P}$ are the scalar and pseudoscalar invariants, respectively. The spacetime metric describing a spherically symmetric, electrically charged AdS ModMax black hole is given by
\begin{equation}\label{eq30}
ds^2=-f(r)dt^2+\frac{1}{f(r)}dr^2+r^2\left(d\theta^2+r^2\sin^2\theta d\phi^2\right)
\end{equation} 
where the metric function can be expressed in the following ways
\begin{equation}\label{eq31}
f(r)=\frac{r^2}{l^2}-\frac{2 M}{r}+\frac{ q^2 e^{-\eta }}{r^2}+1
\end{equation}
We use equation \ref{eq10} for writing the effective potential experienced by a charged test particle in the ModMax AdS background
\begin{equation}\label{eq32}
V_{\text{eff}}=-\frac{1}{m^2}\left[q\frac{ Q e^ {-\eta }}{r_+}-q\frac{Q e^{ -\eta }}{r}\right]^2+f(r)
\end{equation}
We have $\left.\frac{dV_{\mathrm{eff}}}{dr}\right|_{r_+}=4\pi T>0$, so classically the test particle cannot cross the potential barrier. We show the plot of the effective potential against $r/r_+$ in Figure \ref{fig3}.
We find that a charged test particle encounters a potential barrier at a radial location $r = r_2$ before reaching the event horizon at $r = r_+$. As a result, the particle is bounced back from this point and cannot classically reach the horizon, indicating that isentropic absorption is forbidden. Also, as the black hole charge is increased, the height and the width of the effective potential decreases. In order to compute the quantum mechanical probability of the particle crossing the potential barrier, we need to evaluate the Euclidean action given in equation \ref{eq27}. For this particular effective potential, it is not possible to solve the integration analytically. So we perform a numerical computation and show the variation of the Euclidean action with different parameters of the ModMax black hole in Figure \ref{f4}.

We find that the qualitative behaviour of the Euclidean action for the ModMax AdS black hole closely parallels that observed for the NED AdS black hole discussed in the previous subsection. For a fixed value of the black hole charge $Q$, the Euclidean action remains nearly constant as the ModMax parameter $\eta$ is increased, followed by a gradual rise. Beyond a certain value of $\eta$, the action grows very rapidly, leading to a corresponding suppression of the tunnelling probability. This behaviour is reflected in the tunnelling rate through its exponential dependence on the Euclidean action.

For a fixed value of $\eta$, increasing the black hole charge $Q$ leads to a decrease in the Euclidean action, as shown in Figure \ref{f4a}, indicating an enhancement of the tunnelling probability. In this case, the action decreases rapidly with $Q$ before approaching an approximately constant value. Conversely, for fixed value of $Q$, increasing $\eta$ results in an increase of the Euclidean action (Figure \ref{f4b}), implying a suppression of the tunnelling process.

We also examine the dependence of the Euclidean action on the event horizon radius of the black hole. For all choices of parameters, the action exhibits a nonlinear increase with the horizon radius, as illustrated in Figures \ref{f4c} and \ref{f4d}. This behaviour indicates that larger black holes are associated with a stronger suppression of tunnelling, whereas smaller black holes possess a comparatively higher tunnelling probability. As a result, smaller black holes appear less classical in nature and are more susceptible to isentropic absorption processes.

An interesting feature specific to the ModMax system is that increasing the nonlinear parameter $\eta$ causes the Euclidean action to grow more abruptly with the horizon radius, as evident from Figure \ref{f4d}. This suggests that for black holes coupled to stronger nonlinear electrodynamics, the tunnelling probability becomes rapidly suppressed as the black hole size increases.

\subsection{Euler-Heisenberg AdS Black Holes}
\begin{figure}[h!]
\centering
\includegraphics[width=0.5\linewidth]{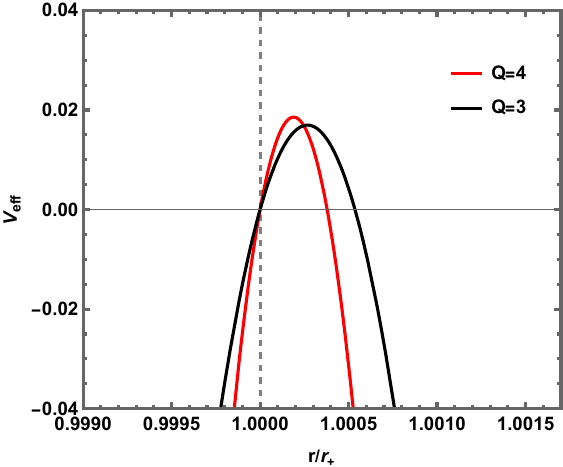}
\caption{Effective potential experienced by the charged test particle for an isentropic absorption by Euler-Heidenberg AdS black hole} 
\label{fig5}
\end{figure}

The third black hole under consideration is the Euler-Heisenberg AdS black hole with the following action \cite{Ple,Sal}
\begin{equation}\label{eq33}
S=\frac{1}{4\pi}\int_{M^4} d^4x \sqrt{-g}\left[\frac{1}{4}(R-2\Lambda)-
\mathcal{L}(F,G) \right]
\end{equation}
here $g$ is again the determinant of the metric tensor, $R$, the Ricci scalar, $\mathcal{L}(F,G)$ denotes the NLED Lagrangian that is related to the electromagnetic invariants as, $F=\frac{1}{4}F_{\mu \nu} F^{\mu \nu}$ and $G=\frac{1}{4}F_{\mu \nu}
{^*F^{\mu \nu}}$ with $F_{\mu\nu}$ denoting the electromagnetic field strength. For the Euler Heisenberg NLED the Lagrangian density is expressed as follows
\begin{equation}\label{eq34}
\mathcal{L}(F,G)=-F+\frac{a}{2}F^2+ \frac{7a}{8} G^2
\end{equation}
where $a$ is the Euler Heisenberg parameter and it determines the strength of the NLED contributions. Also, $a=8\alpha^2/45m^4$, where $\alpha$ is the fine structure constant and $m$ is the mass of the electron; for $a=0$ we recover the Maxwell electrodynamics. The line element of the Euler Heisenberg AdS black hole is given by 
\begin{equation}\label{eq35}
ds^2=f(r)dt^2+\frac{dr^2}{f(r)}+r^2(d\theta^2+\sin^2\theta d\phi^2)
\end{equation}
with the metric function
\begin{equation}\label{eq36}
f(r)= 1-\frac{2M}{r}+\frac{Q^2}{r^2}+\frac{ r^2}{l^2}-\frac{a Q^4}{20 r^6},
\end{equation}
Using equation \ref{eq10} we write down the effective potential experienced by the test charged particle around the Euler Heisenberg black hole
\begin{equation}\label{eq37}
V_{\text{eff}}=-\frac{1}{m^2}\left[q \left(Q-\frac{a Q^3}{10r_{+}^{4}}\right)-q \left(Q-\frac{a Q^3}{10 r^4}\right)\right]^2+f(r)
\end{equation}
We can again easily verify that $\left.\frac{dV_{\mathrm{eff}}}{dr}\right|_{r_+}=4\pi T>0$ and hence, classically, a test particle cannot propagate into this region, meaning that a classical isentropic absorption is prohibited. We show the plot of the effective potential experienced by a charged test particle in Figure \ref{fig5}
In the plot of effective potential against $r/r_+$, the region from $r=r_+$ to the region where $V_\text{eff}$ goes to zero once again (say at $r=r_2$) is the classically forbidden region. A test particle cannot cross the barrier at $r_2$ to reach the horizon and hence an isentropic absorption is also forbidden classically.

\begin{figure}[h!]
    \centering
    \begin{subfigure}[b]{0.45\textwidth}
        \centering
        \includegraphics[width=\textwidth]{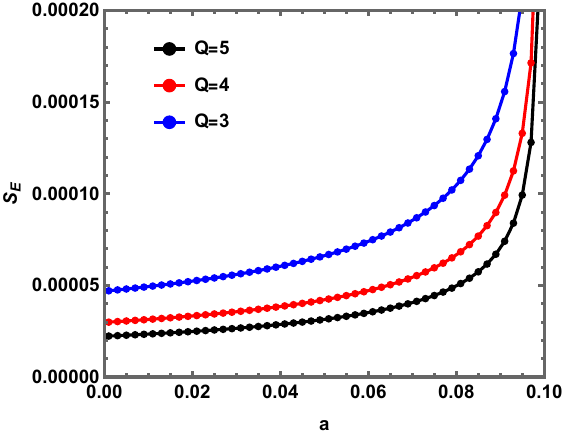}
        \caption{$S_E$ versus $a$ for fixed $Q$}
        \label{f6a}
    \end{subfigure}
    \hspace{0.03\textwidth}
    \begin{subfigure}[b]{0.45\textwidth}
        \centering
        \includegraphics[width=\textwidth, height=0.77\textwidth]{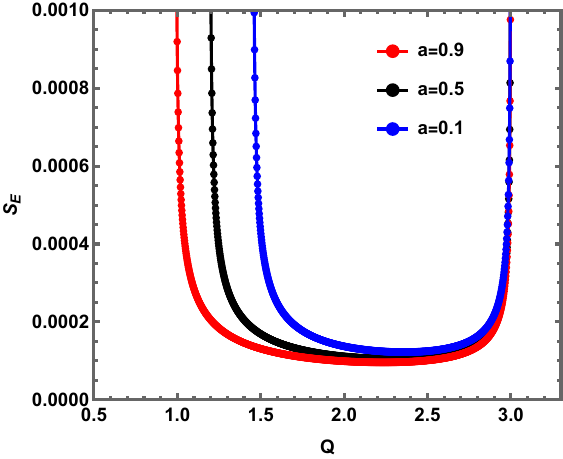}
        \caption{$S_E$ versus $Q$ for fixed $a$}
        \label{f6b}
    \end{subfigure}

    \vskip\baselineskip  

    \begin{subfigure}[b]{0.45\textwidth}
        \centering
        \includegraphics[width=\textwidth]{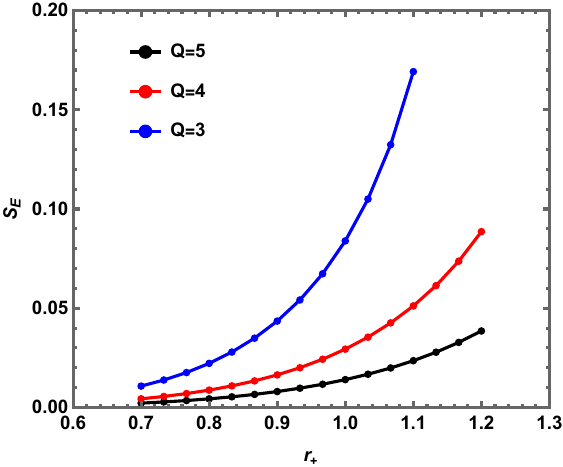}
        \caption{$S_E$ versus $r_+$ for fixed $a$}
        \label{f6c}
    \end{subfigure}
    \hspace{0.03\textwidth}
    \begin{subfigure}[b]{0.45\textwidth}
        \centering
        \includegraphics[width=\textwidth]{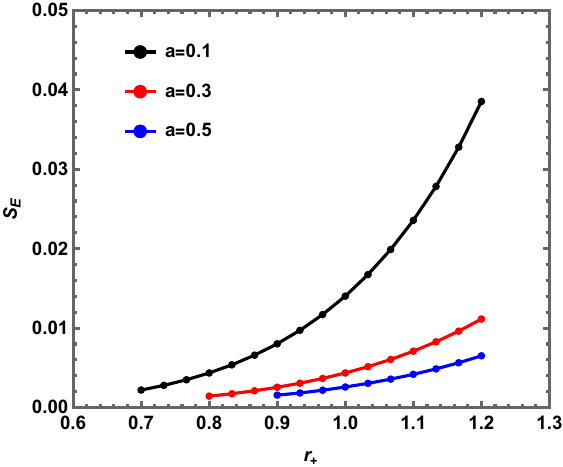}
        \caption{$S_E$ versus $r_+$ for fixed $Q$}
        \label{f6d}
    \end{subfigure}

    \caption{The plots presented above depict the Euclidean action associated with the isentropic absorption of a charged particle by Euler-Heisenberg-AdS black hole. Here we considered $m=0.0002$, $q=0.005$}
    \label{fig6}
\end{figure}

We now compute the quantum mechanical probability of the test charged particle being absorbed isentropically by the Euler-Heisenberg black hole. In order to do that we shall first calculate the Euclidean action as expressed in equation \ref{eq27}. For the effective potential given in equation \ref{eq37}, it is not possible to compute the integral analytically. So we do a numerical integration in the same way as done earlier and plot the Euclidean action with the different black hole parameters. The plots are shown in Figure \ref{fig6}.
The dependence of the Euclidean action on the nonlinear parameter, identified here with the Euler–Heisenberg coupling $a$, exhibits behaviour qualitatively similar to that observed in the previously discussed black hole systems. For a fixed value of the black hole charge $Q$, the Euclidean action initially increases gradually as the Euler-Heisenberg parameter $a$ is raised, but beyond a certain threshold it grows sharply, as illustrated in Figure \ref{f6a}. Owing to the exponential dependence of the tunnelling probability on the Euclidean action, this rapid increase corresponds to a strong suppression of the tunnelling process. Furthermore, for a fixed value of the Euler-Heisenberg parameter $a$, an increase in the black hole charge leads to a decrease in the Euclidean action, thereby enhancing the tunnelling probability.

In contrast to the previously discussed cases, the dependence of the Euclidean action on the black hole charge exhibits a distinctive behaviour for the Euler–Heisenberg system when the nonlinear parameter is held fixed. As shown in Figure \ref{f6b}, for a given value of the Euler-Heisenberg parameter $a$, the Euclidean action initially decreases sharply as the charge $Q$ is increased, leading to a substantial enhancement of the tunnelling probability. Upon further increase of $Q$, the action enters a regime in which it remains approximately constant, before rising abruptly beyond a certain value of charge. This sudden increase results in a strong suppression of the tunnelling probability. A particularly notable feature of the Euler-Heisenberg case is that, for fixed charge $Q$, the Euclidean action decreases with increasing nonlinear parameter $a$, implying an enhancement of the tunnelling probability. This behaviour stands in marked contrast to the black hole systems considered earlier, where increasing the nonlinear parameter led to a increase in the Euclidean action and a corresponding suppression of tunnelling. In the present case, stronger nonlinear effects therefore appear to facilitate isentropic absorption. Another striking aspect is that, for all values of the Euler-Heisenberg parameter $a$, the abrupt rise in the Euclidean action occurs at the same critical value of the black hole charge $Q$. As a consequence, the curves corresponding to different values of $a$ overlap in this region, indicating a universal charge threshold beyond which the tunnelling probability is strongly suppressed.

The dependence of the Euclidean action on the black hole size, characterized by the event horizon radius $r_+$, is shown in Figures \ref{f6c} and \ref{f6d}. For all parameter choices, the Euclidean action increases nonlinearly with the horizon radius, indicating a progressive suppression of the tunnelling probability as the black hole size grows. In contrast, smaller black holes exhibit a comparatively higher probability for isentropic particle absorption, suggesting that they behave in a less classical manner. This qualitative behaviour is consistent with that observed for the previously studied black hole systems. Furthermore, the distinctive feature of the Euler-Heisenberg case, namely, that increasing the nonlinear parameter reduces the Euclidean action for fixed charge (or equivalently, fixed horizon radius) is also evident here, as illustrated in Figure \ref{f6d}.

\subsection{Born-Infeld AdS Black Holes}
\begin{figure}[h!]
\centering
\includegraphics[width=0.5\linewidth]{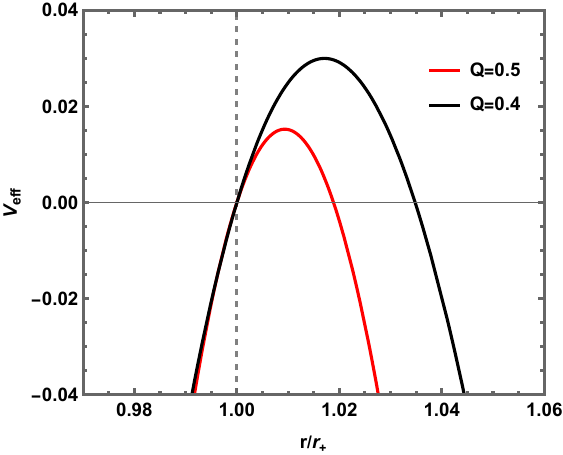}
\caption{Effective potential experienced by the charged test particle for an isentropic absorption by Born-Infeld AdS black hole} 
\label{fig7}
\end{figure}

The last black hole system under consideration is the Born-Infeld AdS system. The four dimensional Born-Infeld action is given by \cite{Born:1934gh}
\begin{equation}\label{eq38}
S=\frac{1}{16\pi}\int{d^4x\sqrt{-g}(R-2\Lambda+\mathcal L(F))}
\end{equation}
where the non linear electromagnetic term is expressed as 
\begin{equation}\label{eq39}
\mathcal L(F)=4\beta^2\left(-\sqrt{1+\frac{F^{\mu\nu}F_{\mu\nu}}{2\beta^2}}\right)
\end{equation}
with $\beta$ being the Born-Infeld parameter. The metric of the four-dimensional Born-Infeld AdS black hole is given by
\begin{equation}\label{eq40}
ds^2=-f(r)dt^2+\frac{dr^2}{f(r)}+r^2(d\theta^2+\sin^2\theta d\phi^2)
\end{equation}
In this case the metric function is given by
\begin{equation}\label{eq41}
f(r)=1-\frac{2M}{r}+\frac{r^2}{l^2}+\frac{2\beta^2r^2}{3}(1-\sqrt{1-z})+\frac{4q^2}{3r^2}\,{_2F_1}\left[\frac{1}{4},\frac{1}{2},\frac{5}{4},z\right]
\end{equation}
Here, $_2F_1[a,b,c,z]$ is the hypergeometric function, with $z=-\frac{Q^2}{\beta^2r^{4}_{+}}$
We can now write down the effective potential experienced by the test charges particle in the Born-Infeld background using equation \ref{eq10}
\begin{equation}\label{eq42}
V_{\text{eff}}=\frac{1}{m^2}\left[q\frac{Q \, _2F_1\left[\frac{1}{4},\frac{1}{2};\frac{5}{4};-\frac{Q^2}{r^4_+ \beta ^2}\right]}{r_+}-q\frac{Q \, _2F_1\left[\frac{1}{4},\frac{1}{2};\frac{5}{4};-\frac{Q^2}{r^4 \beta ^2}\right]}{r}\right]^2+f(r)
\end{equation}
For this effective potential as well, it can be easily checked that $\left.\frac{dV_{\mathrm{eff}}}{dr}\right|_{r_+}=4\pi T>0$. Hence, classically, the particle's motion will be forbidden. We may also observe it from the plot of the effective potential against $r/r_+$ shown in Figure \ref{fig7}, where the particle will experience a potential barrier at say $r=r_2$ before reaching the event horizon at $r_+$, therefore implying that an isentropic absorption is not possible classically.

\begin{figure}[h!]
    \centering
    \begin{subfigure}[b]{0.45\textwidth}
        \centering
        \includegraphics[width=\textwidth]{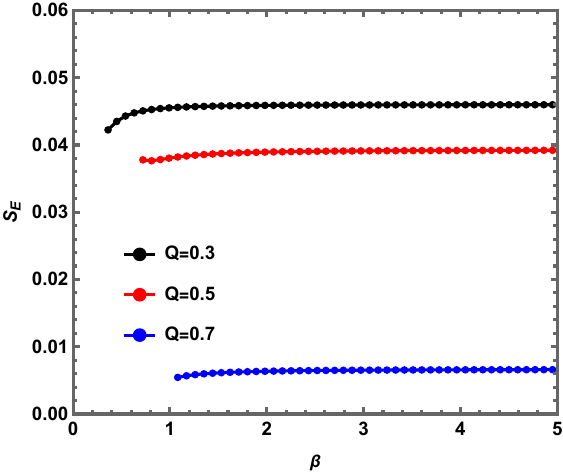}
        \caption{$S_E$ versus $\beta$ for fixed $Q$}
        \label{f8a}
    \end{subfigure}
    \hspace{0.03\textwidth}
    \begin{subfigure}[b]{0.45\textwidth}
        \centering
        \includegraphics[width=\textwidth]{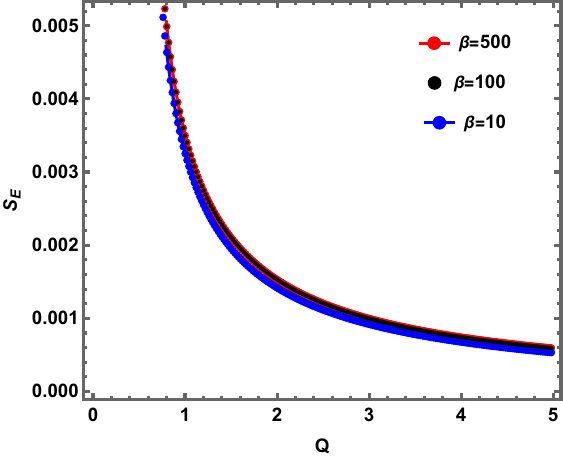}
        \caption{$S_E$ versus $Q$ for fixed $\beta$}
        \label{f8b}
    \end{subfigure}

    \vskip\baselineskip  

    \begin{subfigure}[b]{0.45\textwidth}
        \centering
        \includegraphics[width=\textwidth]{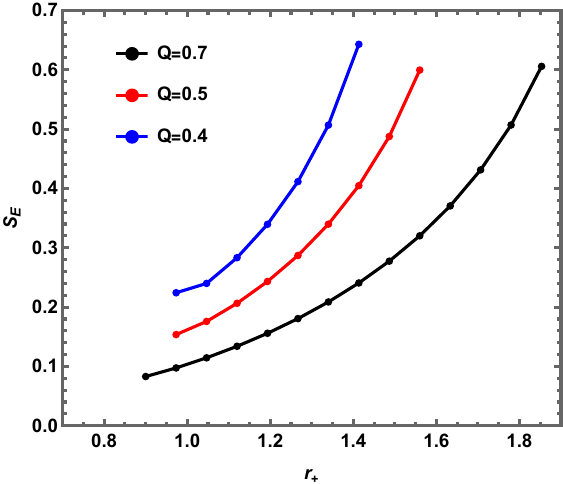}
        \caption{$S_E$ versus $r_+$ for fixed $\beta$}
        \label{f8c}
    \end{subfigure}
    \hspace{0.03\textwidth}
    \begin{subfigure}[b]{0.45\textwidth}
        \centering
        \includegraphics[width=\textwidth]{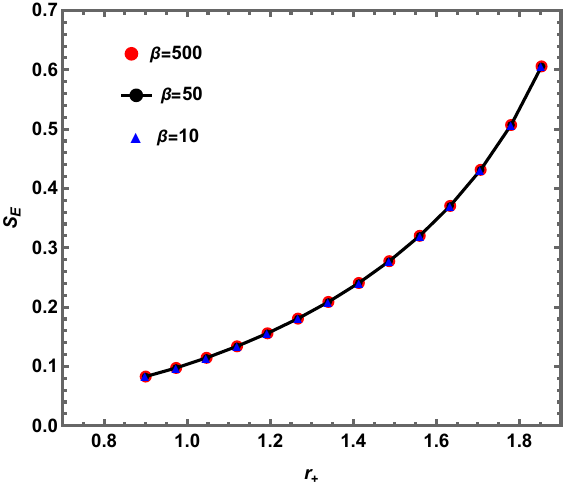}
        \caption{$S_E$ versus $r_+$ for fixed $Q$}
        \label{f8d}
    \end{subfigure}

    \caption{The plots presented above depict the Euclidean action associated with the isentropic absorption of a charged particle by Born-Infeld-AdS black hole.}
    \label{fig8}
\end{figure}

To find the quantum mechanical probability of the test particle to cross the potential barrier we again follow the procedures as earlier. First we compute the Euclidean action in equation \ref{eq37}. It is again not possible to compute the integral analytically and therefore we do it numerically and plot the Euclidean action with the different parameters of the Born-Infeld black hole in Figure \ref{fig8}.
For a fixed value of the black hole charge, we find that the Euclidean action remains nearly constant as the Born-Infeld parameter $\beta$ is increased, as shown in Figure \ref{f8a}. Consequently, variations in the strength of the Born-Infeld nonlinearity have little effect on the tunnelling probability in this regime. However, for a fixed value of $\beta$, the Euclidean action decreases as the charge of the black hole is increased, indicating that the tunnelling probability enhances for black holes with higher charge. This qualitative behaviour is consistent with that observed across all black hole systems examined in our work.

Figure \ref{f8b} further illustrates that the Euclidean action decreases approximately exponentially as the black hole charge increases, leading to a pronounced enhancement of the tunnelling probability. Moreover, for fixed charge $Q$, increasing the Born-Infeld parameter $\beta$ results in an increase of the Euclidean action. Nevertheless, even for large values of $\beta$, this increase remains relatively mild, suggesting that the sensitivity of the tunnelling process to the Born-Infeld nonlinearity is comparatively weak.

Finally, we show the variation of the Euclidean action with the horizon radius $r_+$. For the Born-Infeld black hole also, we find that for all choices of parameters, the Euclidean action increases with the increase in horizon radius as depicted in Figures \ref{f8c} and \ref{f8d}. This indicates that the tunnelling probability for an isentropic absorption is comparatively higher for smaller black holes implying that smaller black holes behave less classically. This result is also consistent across all four NLED black holes considered in our study. Furthermore, Fig.~\ref{f8d} shows that for a fixed value of the horizon radius, variations in the Born–Infeld parameter lead to only marginal changes in the Euclidean action. This observation further reinforces the conclusion that the tunnelling probability exhibits a weak dependence on the nonlinear parameter in Born–Infeld black hole backgrounds.

\section{conclusion and discussion}\label{sec4}
In this work, we have investigated isentropic processes in four AdS black hole systems coupled to nonlinear electrodynamics, namely the NED-AdS, ModMax-AdS, Euler-Heisenberg-AdS, and Born-Infeld-AdS black holes. We demonstrated that, at the classical level, isentropic absorption of a charged test particle is forbidden in all of these backgrounds. For each case, we explicitly identified and illustrated the corresponding classically forbidden region. In particular, we showed that there exists a radial turning point outside the event horizon at which the particle is reflected, preventing it from reaching the horizon through classical motion.

We next examined whether such isentropic processes can occur through quantum mechanical tunnelling. We numerically evaluated the Euclidean action and analyzed its dependence on the relevant black hole parameters. Our results show that, for the NED, ModMax, and Euler-Heisenberg AdS black holes, the Euclidean action increases with the corresponding nonlinear parameters, namely $b$, $\eta$, and $a$, leading to a suppression of the tunnelling probability as the strength of the nonlinearity is enhanced. For the Born-Infeld case, the Euclidean action exhibits only a weak dependence on the nonlinear parameter: after an initial increase, it becomes almost constant, resulting in a tunnelling probability that remains nearly unchanged. Overall, these findings indicate that isentropic absorption via tunnelling becomes progressively less probable as nonlinear electrodynamic effects become more pronounced.

We find that the Euclidean action generally decreases with increasing black hole charge, indicating that isentropic absorption via tunnelling is relatively more probable for black holes with larger charge. An exception occurs in the Euler-Heisenberg case, where this trend holds only up to a certain value of the charge $Q$. Beyond this point, the Euclidean action increases, leading to a corresponding suppression of the tunnelling probability at higher charge. This behaviour is also reflected in the plot of the Euclidean action against the horizon radius.

We further find that the tunnelling probability decreases with increasing black hole size, irrespective of the specific choice of parameters. This suggests that smaller black holes exhibit less classical behaviour, as they are more prone to isentropic absorption of charged particles and can more easily give rise to violations of entropy bounds compared to larger black holes. A similar trend was also reported in Ref. \cite{Dubey:2025imq}.These observations suggest that this behaviour may represent a universal feature of all black hole systems.

Our findings support the perspective that entropy bounds are firmly upheld within a semiclassical description, while allowing for the possibility of well-defined nonperturbative deviations in black hole settings. Such nonperturbative processes are typically highly suppressed and therefore remain absent in standard semiclassical analyses, which primarily account for perturbative or thermal quantum corrections. As a result, entropy-bound relations established within the semiclassical framework continue to remain valid, and it is reasonable to neglect these nonperturbative contributions when one considers observables averaged over all possible quantum processes. In this sense, nonperturbative effects may provide an important ingredient in developing a consistent understanding of black hole dynamics and could offer insights into the resolution of the black hole information paradox \cite{Ong:2016iwi,Raju:2020smc}.

We conclude by clarifying the precise scope of our results.

First, what has been computed in this work is the semiclassical tunnelling amplitude for a charged test particle to be absorbed under the isentropic condition $dS=0$. The calculation is performed in the fixed-background approximation, neglecting backreaction and self-force effects.

Second, from this amplitude we infer the existence of a nonperturbative channel through which classically forbidden isentropic absorption may occur. This channel is suppressed but nonzero for non-extremal black holes.

Third, the connection to entropy bounds should be interpreted with care. The entropy appearing in the isentropic condition is the Bekenstein-Hawking entropy, $S_{BH} = \frac{A}{4}$,
defined geometrically in terms of the horizon area. We do not compute a microscopic (Boltzmann) entropy or count underlying degrees of freedom.

Therefore, our results do not demonstrate a violation of entropy bounds in a strict statistical sense. Rather, they suggest that semiclassical tunnelling may allow processes that evade classical entropy-increase arguments. Whether such channels modify entropy bounds beyond the test-particle approximation requires inclusion of backreaction and remains an open question.

The universality suggested by the behaviour across different nonlinear electrodynamics models should thus be regarded as a structural feature of the semiclassical effective potential, rather than a proof of fundamental entropy-bound violation.

It would be worthwhile to extend the present analysis to rotating and charged black holes in higher dimensions, where additional geometric and dynamical features may significantly influence the isentropic absorption process. Another natural direction is to investigate quantum-corrected black hole solutions and examine how such corrections modify the associated Euclidean action and, consequently, the tunnelling probability. Furthermore, it remains an open question how island contributions \cite{Almheiri:2019hni} behave when isentropic processes accumulate over the lifetime of a black hole, and whether the inclusion of such contributions destabilizes or invalidates the underlying semiclassical background \cite{Marolf:2020rpm}. These issues merit detailed investigation and are left for future work.

\appendix
\section{Gauge Potential, Thermodynamic Potential, and Isentropic Condition}

In this Appendix we clarify the relation between the thermodynamic electrostatic potential $\Phi$, the gauge potential $A_t(r)$ appearing in the particle dynamics, and the effective potential used in the tunnelling analysis. This addresses potential gauge ambiguities in nonlinear electrodynamics (NLED) models.

\subsection{Gauge potential and thermodynamic potential}

For a static, spherically symmetric charged black hole, the metric takes the form
\begin{equation}
ds^2 = - f(r)\, dt^2 + \frac{dr^2}{f(r)} + r^2 d\Omega^2.
\end{equation}

A charged test particle of mass $m$ and charge $q$ moving radially in this background is governed by the Lagrangian
\begin{equation}
\mathcal{L} = -m \sqrt{-g_{\mu\nu} \dot{x}^\mu \dot{x}^\nu} + q A_\mu \dot{x}^\mu.
\end{equation}

For purely radial motion, the conserved energy is
\begin{equation}
E = - p_t = m f(r) \dot{t} - q A_t(r).
\end{equation}

The radial equation of motion can be written in the standard form
\begin{equation}
\left( \frac{dr}{d\tau} \right)^2 + V_{\rm eff}(r) = 0,
\end{equation}
with effective potential
\begin{equation}
\label{Veff_full}
V_{\rm eff}(r)
=
- \frac{1}{m^2}\big( E - q A_t(r) \big)^2
+ f(r).
\end{equation}

The physical electrostatic potential entering the first law of black hole thermodynamics is defined as the potential difference between infinity and the event horizon,
\begin{equation}
\Phi = A_t(\infty) - A_t(r_+).
\end{equation}

The electrostatic potential conjugate to the charge is defined thermodynamically via the first law,
\begin{equation}
dM = T dS + \Phi\, dQ,
\end{equation}
so that
\begin{equation}
\Phi = \left( \frac{\partial M}{\partial Q} \right)_S .
\end{equation}

For the NED-AdS black hole considered in the main text, differentiating the mass function with respect to $Q$ at fixed horizon radius $r_+$ yields
\begin{equation}
\Phi
=
\frac{Q}{b}
\tanh\!\left( \frac{b}{r_+} \right).
\end{equation}

On the other hand, the gauge potential entering the Lorentz force term is
\begin{equation}
A_t(r)
=
\frac{Q}{b}
\tanh\!\left( \frac{b}{r} \right).
\end{equation}

Since $\tanh(b/r) \to 0$ as $r \to \infty$, we have
\begin{equation}
A_t(\infty) = 0.
\end{equation}

Therefore,
\begin{equation}
\Phi = A_t(\infty) - A_t(r_+) = -A_t(r_+).
\end{equation}

With the sign convention used in the particle Lagrangian, the conserved energy appears in the combination $E - qA_t(r)$. Hence the isentropic condition
\begin{equation}
E - q\Phi = 0
\end{equation}
reduces to
\begin{equation}
E - qA_t(r_+) = 0.
\end{equation}

This shows explicitly that the thermodynamic potential coincides with the horizon value of the gauge potential (up to the sign convention fixed by the potential difference).

An analogous relation holds for the other nonlinear electrodynamics models considered in this work, where, in each case, the thermodynamic potential $\Phi = (\partial M/\partial Q)_S$ coincides with the horizon value of $A_t(r)$ under the standard gauge choice $A_t(\infty)=0$.

\subsection{Effective potential under the isentropic condition}

Identifying $dM \to E$ and $dQ \to q$ for particle absorption, the isentropic condition becomes
\begin{equation}
E - q \Phi = 0.
\end{equation}

Using $\Phi = -A_t(r_+)$ and the sign convention above, this gives
\begin{equation}
E - q A_t(r_+) = 0.
\end{equation}

Substituting this into the full effective potential
(\ref{Veff_full}) yields
\begin{align}
V_{\rm eff}(r)
&=
- \frac{1}{m^2}
\big( q A_t(r_+) - q A_t(r) \big)^2
+ f(r)
\\
&=
- \frac{q^2}{m^2}
\left[ A_t(r_+) - A_t(r) \right]^2
+ f(r).
\end{align}

This is the effective potential used in the tunnelling analysis in the main text.

\subsection{Gauge invariance}

Under a constant gauge shift
\begin{equation}
A_t(r) \;\longrightarrow\; A_t(r) + C,
\end{equation}
the conserved energy transforms as
\begin{equation}
E \;\longrightarrow\; E + q C.
\end{equation}

Therefore, the combination
\begin{equation}
E - q A_t(r)
\end{equation}
is gauge invariant.

Consequently, the effective potential $V_{\rm eff}(r)$ and the tunnelling probability computed from it are manifestly gauge independent.

This establishes that the isentropic tunnelling analysis presented in the main text is consistent and free of gauge ambiguities.

\section{Near-Horizon Behaviour of the Effective Potential}

In this Appendix we explicitly demonstrate that, under the isentropic condition, the effective potential vanishes at the horizon and has a positive radial derivative there. This establishes the existence of a classically forbidden region outside the event horizon for any non-extremal static black hole of the form considered in the main text.

We begin with the effective potential
\begin{equation}
V_{\rm eff}(r)
=
- \frac{1}{m^2}\big( E - qA_t(r) \big)^2
+ f(r),
\end{equation}
where $f(r_+)=0$ defines the event horizon.

\subsection{Vanishing of $V_{\rm eff}$ at the horizon}

For an isentropic process,
\begin{equation}
E - q\Phi = 0,
\end{equation}
and, as shown in Appendix A,
\begin{equation}
\Phi = A_t(\infty) - A_t(r_+) = -A_t(r_+),
\end{equation}
so that
\begin{equation}
E - qA_t(r_+) = 0.
\end{equation}

Since $f(r_+)=0$, it follows immediately that
\begin{equation}
V_{\rm eff}(r_+) = 0.
\end{equation}

\subsection{Radial derivative at the horizon}

Differentiating the effective potential with respect to $r$ gives
\begin{equation}
\frac{dV_{\rm eff}}{dr}
=
\frac{2q}{m^2}
\big( E - qA_t(r) \big) A_t'(r)
+
f'(r).
\end{equation}

Evaluating at $r=r_+$ and using the isentropic condition
\begin{equation}
E - qA_t(r_+) = 0,
\end{equation}
we obtain
\begin{equation}
\left.\frac{dV_{\rm eff}}{dr}\right|_{r_+}
=
f'(r_+).
\end{equation}

Thus the contribution from the electromagnetic sector vanishes at the horizon, and the derivative is entirely determined by the geometry.

For a static black hole with metric
\begin{equation}
ds^2 = -f(r) dt^2 + \frac{dr^2}{f(r)} + r^2 d\Omega^2,
\end{equation}
the Hawking temperature is given by
\begin{equation}
T = \frac{f'(r_+)}{4\pi}.
\end{equation}

Therefore,
\begin{equation}
\left.\frac{dV_{\rm eff}}{dr}\right|_{r_+}
=
4\pi T.
\end{equation}

For any non-extremal black hole, $T>0$, and hence
\begin{equation}
\left.\frac{dV_{\rm eff}}{dr}\right|_{r_+} > 0.
\end{equation}

Consequently, near the horizon,
\begin{equation}
V_{\rm eff}(r)
=
4\pi T (r-r_+)
+
\mathcal{O}((r-r_+)^2),
\end{equation}
which is positive for $r>r_+$ sufficiently close to the horizon. This establishes that the isentropic absorption process is classically forbidden for all non-extremal static black holes considered in this work.

\section{Derivation of the Euclidean Action from the Radial Equation}

In this Appendix we provide a detailed derivation of the Euclidean action used in the tunnelling analysis, explicitly connecting the radial equation of motion to the standard WKB expression.

\subsection{Effective one-dimensional Lagrangian}

For a charged particle of mass $m$ moving radially in a static background
\begin{equation}
ds^2 = -f(r) dt^2 + \frac{dr^2}{f(r)} + r^2 d\Omega^2,
\end{equation}
the worldline action is \cite{Wald}
\begin{equation}
S = -m \int d\tau + q \int A_\mu dx^\mu.
\end{equation}

Eliminating $t$ in favour of the conserved energy $E$, the radial motion reduces to an effective one-dimensional system governed by
\begin{equation}
\left( \frac{dr}{d\tau} \right)^2 + V_{\rm eff}(r) = 0,
\end{equation}
where
\begin{equation}
V_{\rm eff}(r)
=
- \frac{1}{m^2}(E - qA_t(r))^2 + f(r).
\end{equation}

This equation plays the role of a Hamiltonian constraint for a one-dimensional system with zero total energy.

\subsection{Canonical momentum and WKB form}

The canonical radial momentum is
\begin{equation}
p_r = m \frac{dr}{d\tau}.
\end{equation}

Using the radial equation,
\begin{equation}
p_r^2 = m^2 \left( \frac{dr}{d\tau} \right)^2
= - m^2 V_{\rm eff}(r).
\end{equation}

In the classically forbidden region, $V_{\rm eff}(r) > 0$, so that
\begin{equation}
p_r = \pm i m \sqrt{V_{\rm eff}(r)}.
\end{equation}

The standard WKB tunnelling exponent is
\begin{equation}
S_E = \int |p_r| \, dr.
\end{equation}

Therefore,
\begin{equation}
S_E = m \int_{r_+}^{r_2} \sqrt{V_{\rm eff}(r)} \, dr.
\end{equation}

\subsection{Relation to the Euclidean proper-time derivation}

Alternatively, starting from
\begin{equation}
\left( \frac{dr}{d\tau} \right)^2 = -V_{\rm eff}(r),
\end{equation}
we perform a Wick rotation
\begin{equation}
\tau \rightarrow -i \tau_E,
\end{equation}
giving
\begin{equation}
\left( \frac{dr}{d\tau_E} \right)^2 = V_{\rm eff}(r).
\end{equation}

Thus,
\begin{equation}
d\tau_E = \frac{dr}{\sqrt{V_{\rm eff}(r)}}.
\end{equation}

The Euclidean action becomes
\begin{equation}
S_E = m \int d\tau_E
= m \int_{r_+}^{r_2}
\frac{dr}{\sqrt{V_{\rm eff}(r)}}.
\end{equation}

The two expressions are equivalent upon using
\begin{equation}
p_r = m \frac{dr}{d\tau_E}
= m \sqrt{V_{\rm eff}(r)}.
\end{equation}

The precise normalization depends on whether one works in proper-time or Hamiltonian form, but both yield the same exponential tunneling factor.

\subsection{Tunneling probability}

In the saddle-point approximation, the tunnelling probability is
\begin{equation}
\Gamma \sim e^{-2 S_E}.
\end{equation}

This establishes explicitly the connection between the radial equation and the WKB tunnelling exponent used in the main text.

\section{Numerical Implementation in \textit{Mathematica}}

All numerical computations were performed using \textit{Mathematica}.

\subsection{Determination of the turning point}

For fixed black hole and particle parameters, the outer turning point $r_2$ is defined by
\begin{equation}
V_{\rm eff}(r_2) = 0,
\qquad r_2 > r_+.
\end{equation}

The root was obtained using the in-built \textit{Mathematica} function \texttt{FindRoot}, with initial bracketing determined by scanning the sign change of $V_{\rm eff}(r)$ outside the horizon from its plot.

\subsection{Treatment of near-horizon behaviour}

Near the horizon, $V_{\rm eff}(r)$ becomes complicated for the NLED systems considered, and so the integrand becomes very heavy to handle. Although, it is possible to perform the integral, it becomes numerically stiff.

To control this behaviour, we used \texttt{NIntegrate} with the options
\begin{verbatim}
Method -> {"GlobalAdaptive","SingularityDepth"->10}, MaxRecursion -> 20
\end{verbatim}

In addition, convergence was tested by increasing \texttt{MaxRecursion} up to 30, increasing the
\texttt{WorkingPrecision} to 30 digits, both resulted values were stable under these variations to at least six significant digits.

\subsection{Integration routine and Parameter Dependence}

The Euclidean action was computed as
\begin{equation}
S_E = \int_{r_+}^{r_2} \frac{dr}{\sqrt{V_{\rm eff}(r)}},
\end{equation}
using \texttt{NIntegrate} with adaptive mesh refinement. Stability was verified by varying tolerances and step parameters.

In representative plots, the particle parameters were fixed to $m=0.0002$ and $q=0.005$. Since these corresponds to the charged test particle, the mass and charge of the same is considered to be much much smaller than those of the black hole.

Varying these parameters within an order of magnitude modifies the overall scale of $S_E$ but does not alter the existence of the turning point, the monotonic dependence on black hole parameters and the qualitative trends shown in the figures. The conclusions of the paper are therefore insensitive to moderate changes in the particle parameters.


\begin{thebibliography}{99}


\bibitem{Hawking}
S.~W.~Hawking,
``Particle Creation by Black Holes,''
Commun. Math. Phys. \textbf{43} (1975), 199-220
[erratum: Commun. Math. Phys. \textbf{46} (1976), 206]
doi:10.1007/BF02345020

\bibitem{Hawking2}
S.~W.~Hawking,
``Breakdown of Predictability in Gravitational Collapse,''
Phys. Rev. D \textbf{14}, 2460-2473 (1976).

\bibitem{Yeom}
D.~Yeom and H.~Zoe,
``Semi-classical black holes with large N re-scaling and information loss problem,''
Int. J. Mod. Phys. A \textbf{26}, 3287-3314 (2011)
[arXiv:0907.0677 [hep-th]].

\bibitem{Almheiri:2012rt}
A.~Almheiri, D.~Marolf, J.~Polchinski and J.~Sully,
``Black Holes: Complementarity or Firewalls?,''
JHEP \textbf{02}, 062 (2013)
[arXiv:1207.3123 [hep-th]].

\bibitem{Maldacena:1997re}
J.~M.~Maldacena,
``The Large $N$ limit of superconformal field theories and supergravity,''
Adv. Theor. Math. Phys. \textbf{2}, 231-252 (1998)
[arXiv:hep-th/9711200 [hep-th]].

\bibitem{Banks:1983by}
T.~Banks, L.~Susskind and M.~E.~Peskin,
``Difficulties for the Evolution of Pure States Into Mixed States,''
Nucl. Phys. B \textbf{244}, 125-134 (1984).

\bibitem{Ong:2016iwi}
Y.~C.~Ong and D.~Yeom,
``Summary of Parallel Session: Black Hole Evaporation and Information Loss Paradox,''
Proceedings of the 2nd LeCosPA Symposium
[arXiv:1602.06600 [hep-th]].

\bibitem{Bekenstein}
J.~D.~Bekenstein,
``Black holes and entropy,''
Phys. Rev. D \textbf{7}, 2333-2346 (1973).

\bibitem{Iyer}
V.~Iyer and R.~M.~Wald,
``Some properties of Noether charge and a proposal for dynamical black hole entropy,''
Phys. Rev. D \textbf{50}, 846-864 (1994)
[arXiv:gr-qc/9403028 [gr-qc]].


\bibitem{Bekenstein:1974ax}
J.~D.~Bekenstein,
``Generalized second law of thermodynamics in black hole physics,''
Phys. Rev. D \textbf{9}, 3292-3300 (1974)
doi:10.1103/PhysRevD.9.3292

\bibitem{Bekenstein:1973ur}
J.~D.~Bekenstein,
``Black holes and entropy,''
Phys. Rev. D \textbf{7} (1973), 2333-2346
doi:10.1103/PhysRevD.7.2333

\bibitem{Bekenstein:1972tm}
J.~D.~Bekenstein,
``Black holes and the second law,''
Lett. Nuovo Cim. \textbf{4}, 737-740 (1972)
doi:10.1007/BF02757029

\bibitem{Bekenstein:1980jp}
J.~D.~Bekenstein,
``A Universal Upper Bound on the Entropy to Energy Ratio for Bounded Systems,''
Phys. Rev. D \textbf{23} (1981), 287
doi:10.1103/PhysRevD.23.287

\bibitem{bound1}
P.~Hayden and J.~Wang,
``What exactly does Bekenstein bound?,''
Quantum \textbf{9} (2025), 1664
doi:10.22331/q-2025-03-20-1664
[arXiv:2309.07436 [hep-th]].

\bibitem{bound2}
R.~Bousso,
``Bound states and the Bekenstein bound,''
JHEP \textbf{02} (2004), 025
doi:10.1088/1126-6708/2004/02/025
[arXiv:hep-th/0310148 [hep-th]].

\bibitem{bound3}
A.~Pesci,
``A proof of the Bekenstein bound for any strength of gravity through holography,''
Class. Quant. Grav. \textbf{27} (2010), 165006
doi:10.1088/0264-9381/27/16/165006
[arXiv:0903.0319 [gr-qc]].

\bibitem{bound4}
R.~Brustein and G.~Veneziano,
``A Causal entropy bound,''
Phys. Rev. Lett. \textbf{84} (2000), 5695-5698
doi:10.1103/PhysRevLett.84.5695
[arXiv:hep-th/9912055 [hep-th]].
\bibitem{bound5}
W.~G.~Unruh and R.~M.~Wald,
``Acceleration Radiation and Generalized Second Law of Thermodynamics,''
Phys. Rev. D \textbf{25} (1982), 942-958
doi:10.1103/PhysRevD.25.942
\bibitem{bound6}
W.~G.~Unruh and R.~M.~Wald,
``Entropy bounds, acceleration radiation, and the generalized second law''
Phys. Rev. D \textbf{27} (1983), 2271-2276
doi:10.1103/PhysRevD.27.2271
\bibitem{bound7}
L.~Buoninfante, G.~G.~Luciano, L.~Petruzziello and F.~Scardigli,
``Bekenstein bound and uncertainty relations,''
Phys. Lett. B \textbf{824} (2022), 136818
doi:10.1016/j.physletb.2021.136818
[arXiv:2009.12530 [hep-th]].
\bibitem{bound8}
T.~Matsuda and S.~Mukohyama,
``Covariant entropy bound beyond general relativity,''
Phys. Rev. D \textbf{103} (2021) no.2, 024002
doi:10.1103/PhysRevD.103.024002
[arXiv:2007.14015 [hep-th]].
\bibitem{bound9}
R.~Bousso,
``A Covariant entropy conjecture,''
JHEP \textbf{07} (1999), 004
doi:10.1088/1126-6708/1999/07/004
[arXiv:hep-th/9905177 [hep-th]].
\bibitem{bound10}
H.~Zhu and J.~Jiang,
``Generalized covariant entropy bound in Einstein gravity with quadratic curvature corrections,''
JHEP \textbf{05} (2024), 286
doi:10.1007/JHEP05(2024)286
[arXiv:2311.05352 [gr-qc]].
\bibitem{bound11}
E.~E.~Flanagan, D.~Marolf and R.~M.~Wald,
``Proof of classical versions of the Bousso entropy bound and of the generalized second law,''
Phys. Rev. D \textbf{62} (2000), 084035
doi:10.1103/PhysRevD.62.084035
[arXiv:hep-th/9908070 [hep-th]].
\bibitem{bound12}
G.~Acquaviva, A.~Iorio and L.~Smaldone,
``Bekenstein bound from the Pauli principle: a brief introduction,''
PoS \textbf{ICHEP2020}, 681 (2021)
doi:10.22323/1.390.0681
[arXiv:2011.05176 [hep-th]].
\bibitem{bound13}
S.~Sachdev,
``Bekenstein-Hawking Entropy and Strange Metals,''
Phys. Rev. X \textbf{5}, no.4, 041025 (2015)
doi:10.1103/PhysRevX.5.041025
[arXiv:1506.05111 [hep-th]].
\bibitem{bound14}
J.~Maldacena and D.~Stanford,
``Remarks on the Sachdev-Ye-Kitaev model,''
Phys. Rev. D \textbf{94}, no.10, 106002 (2016)
doi:10.1103/PhysRevD.94.106002
[arXiv:1604.07818 [hep-th]].
\bibitem{bound15}
G.~Acquaviva, A.~Iorio and M.~Scholtz,
``Quasiparticle picture from the Bekenstein bound,''
PoS \textbf{CORFU2017}, 206 (2017)
doi:10.22323/1.318.0206
[arXiv:1712.05275 [hep-th]].
\bibitem{bound16}
J.~D.~Bekenstein,
``Entropy bounds and black hole remnants,''
Phys. Rev. D \textbf{49}, 1912-1921 (1994)
doi:10.1103/PhysRevD.49.1912
[arXiv:gr-qc/9307035 [gr-qc]].


\bibitem{Mann:2025ojd}
R.~B.~Mann and D.~h.~Yeom,
``Isentropic process of Reissner-Nordstr{\"o}m black holes: A possible excess of the entropy bound via a nonperturbative channel,''
Phys. Rev. D \textbf{112}, no.6, 064042 (2025)
doi:10.1103/t9px-txyf
[arXiv:2505.01663 [gr-qc]].
\bibitem{Dubey:2025imq}
N.~K.~Dubey and S.~Kolekar,
``Isentropic processes for axisymmetric black holes,''
Phys. Rev. D \textbf{112}, no.8, 084018 (2025)
doi:10.1103/731x-38lt
[arXiv:2507.08547 [gr-qc]].


\bibitem{Kruglov:2022mde}
S.~I.~Kruglov,
``NED-AdS black holes, extended phase space thermodynamics and Joule\textendash{}Thomson expansion,''
Nucl. Phys. B \textbf{984}, 115949 (2022)
doi:10.1016/j.nuclphysb.2022.115949
[arXiv:2209.10524 [physics.gen-ph]].

\bibitem{Bron}
K.~A.~Bronnikov,
``Regular magnetic black holes and monopoles from nonlinear electrodynamics,''
Phys. Rev. D \textbf{63}, 044005 (2001)
doi:10.1103/PhysRevD.63.044005
[arXiv:gr-qc/0006014 [gr-qc]].


\bibitem{Sekhmani:2025kav}
Y.~Sekhmani, S.~K.~Maurya, M.~K.~Jasim, {\.I}.~Sakall{\i}, J.~Rayimbaev and I.~Ibragimov,
``Thermodynamics and phase transition of anti de Sitter black holes with ModMax nonlinear electrodynamics and perfect fluid dark matter,''
Eur. Phys. J. C \textbf{85} (2025) no.3, 229
doi:10.1140/epjc/s10052-025-13932-5

\bibitem{Ple}
J.~Plebanski,
``LECTURES ON NON LINEAR ELECTRODYNAMICS,''
RX-476.
\bibitem{Sal}
I.~H.~Salazar, A.~Garcia and J.~Plebanski,
``Duality Rotations and Type $D$ Solutions to Einstein Equations With Nonlinear Electromagnetic Sources,''
J. Math. Phys. \textbf{28} (1987), 2171-2181
doi:10.1063/1.527430

\bibitem{Born:1934gh}
M.~Born and L.~Infeld,
``Foundations of the new field theory,''
Proc. Roy. Soc. Lond. A \textbf{144} (1934) no.852, 425-451
doi:10.1098/rspa.1934.0059

\bibitem{Raju:2020smc}
S.~Raju,
``Lessons from the information paradox,''
Phys. Rept. \textbf{943} (2022), 1-80
doi:10.1016/j.physrep.2021.10.001
[arXiv:2012.05770 [hep-th]].

\bibitem{Almheiri:2019hni}
A.~Almheiri, R.~Mahajan, J.~Maldacena and Y.~Zhao,
``The Page curve of Hawking radiation from semiclassical geometry,''
JHEP \textbf{03} (2020), 149
doi:10.1007/JHEP03(2020)149
[arXiv:1908.10996 [hep-th]].
\bibitem{Marolf:2020rpm}
D.~Marolf and H.~Maxfield,
``Observations of Hawking radiation: the Page curve and baby universes,''
JHEP \textbf{04} (2021), 272
doi:10.1007/JHEP04(2021)272
[arXiv:2010.06602 [hep-th]].

\bibitem{Wald}
R.~M.~Wald,
``General Relativity,''
Chicago Univ. Pr., 1984,
doi:10.7208/chicago/9780226870373.001.0001








\end{thebibliography}
\end{document}